\documentclass[prd,preprint,showpacs]{revtex4}
\usepackage{amsmath}
\usepackage{latexsym}
\usepackage{amsfonts}
\usepackage{graphicx}

\newcommand{\omm}{\Omega_m}
\newcommand{\omk}{\Omega_k}

\begin{document}

\title{Growth factor parametrization in curved space}
\author{Yungui Gong}
\email{gongyg@cqupt.edu.cn}
\affiliation{College of Mathematics and Physics, Chongqing
University of Posts and Telecommunications, Chongqing 400065, China}
\affiliation{Kavli Institute for Theoretical Physics China, CAS,
Beijing 100190, China}

\author{Mustapha Ishak}
\email{mishak@utdallas.edu }
\affiliation{Department of Physics,
The University of Texas at Dallas, Richardson, Texas 75083, USA}

\author{Anzhong Wang}
\email{anzhong_wang@baylor.edu} \affiliation{CASPER, Physics
Department, Baylor University, Waco, Texas 76798, USA}

\begin{abstract}
The growth rate of matter perturbation and the expansion rate of the
Universe can be used to distinguish modified gravity and dark energy
models in explaining cosmic acceleration. We explore here the
inclusion of spatial curvature into the growth factor. We expand
previous results using the approximation $\Omega_{m}^\gamma$ and
then suggest a new form, $f_a=\Omega_m^\gamma+(\gamma-4/7)\Omega_k$, as
an approximation for the growth factor when the curvature $\Omega_k$
is not negligible, and where the growth index $\gamma$ is usually
model dependent. The expression recovers the standard results for
the curved and flat $\Lambda$CDM and Dvali-Gabadadze-Porrati models.
Using the best fit values of $\Omega_{m0}$ and $\Omega_{k0}$ to the
expansion/distance measurements from Type Ia supernovae, baryon
acoustic oscillation, WMAP5, and $H(z)$ data, we fit
the growth index parameter to current growth factor data and obtain
$\gamma_{\Lambda}\left(\Omega_{k} \not= 0\right) =
0.65^{+0.17}_{-0.15}$ and $\gamma_{DGP}\left(\Omega_{k} \not=
0\right) = 0.53^{+0.14}_{-0.12}$. For the $\Lambda$CDM model, the 1-$\sigma$
observational bounds are found consistent with theoretical value, unlike
the case for the Dvali-Gabadadze-Porrati model. We also find that the current data we used is not enough
to put significant constraints when the 3 parameters in $f_a$ are
 fit simultaneously. Importantly, we find that, in the presence of curvature,
the analytical expression proposed for $f_a$ provides a better fit to
the growth factor than other forms and should be useful for future
high precision missions and studies.
\end{abstract}

\pacs{95.36.+x; 98.80.Es; 04.50.-h}
\preprint{CAS-KITPC/ITP-092}
\maketitle
\section{Introduction}

The discovery of late time cosmic acceleration \cite{acc1}
challenges our understanding of the standard models of gravity and
particle physics. Within the framework of Friedmann-Robertson-Walker
cosmology, a dark energy component with negative pressure
 is invoked to explain the observed accelerated
expansion of the Universe. One simple candidate of dark energy which
is consistent with current observations is the cosmological
constant. By choosing a suitable equation of state $w=p/\rho$ for
dark energy, we can recover the observed expansion rate $H(z)$ and
the luminosity distance redshift relation $d_L(z)$. Many parametric
and nonparametric model-independent methods were proposed to study
the property of dark energy, see for example
\cite{astier01,huterer,par2,par3,lind,alam,jbp,par4,par1,
par5,gong04,gong05,gong06,gong08,Ishak2007,rev7,sahni} and references
therein.

The apparent acceleration of the Universe may be explained by the
modification of gravitation, such as for example the
Dvali-Gabadadze-Porrati (DGP) brane-world model \cite{dgp}, in which
gravity appears four dimensional at short distances but is modified
at large distances. Recently, it was shown that the late cosmic
acceleration of the Universe can also be realized in the
Horava-Witten heterotic $M$ theory \cite{MTheory} and string theory
\cite{string} on $S^{1}/Z_{2}$. It is remarkable that the
acceleration is transient in all these models. To distinguish the
effect of modified gravity from dark energy, we can use the growth
rate of large scale in the Universe in addition to the distance
data. While different models give the same late time accelerated
expansion, the growth of matter perturbation they produce differs
\cite{jetp}. Recently, the use of the growth rate of matter
perturbation in addition to the expansion history of the Universe to
differentiate dark energy models and modified gravity attracted much
attention, see, for example, an incomplete list
\cite{huterer07,sereno,knox,ishak,yun,Aquavivaetal,Danieletal, %
polarski,sapone,balles,%
gannouji,berts,laszlo,kunz,kiakotou,porto,wei,ness} and references
therein. To the linear order of perturbation, at large scales, the matter
density perturbation $\delta=\delta\rho_m/\rho_m$ satisfies the
simple equation:
\begin{equation}
\label{denpert}
\ddot{\delta}+2H\dot{\delta}-4\pi G_{\rm eff}\,\rho_m\delta=0,
\end{equation}
where $\rho_m$ is the matter energy density and $G_{\rm eff}$ denotes
the effect of modified gravity. For example,
$G_{\rm eff}/G=(4+2\omega)/(3+2\omega)$ for the Brans-Dicke theory
\cite{boiss} and $G_{\rm eff}/G=1+[3-6r_c H(1+\dot{H}/3H^2)]^{-1}$ for
the DGP model \cite{lue,maartens}, the dimensionless matter energy
density $\omm=8\pi G\rho_m/(3H^2)$. In terms of the growth factor
$f=d\ln\delta/d\ln a$, the matter density perturbation Eq.
(\ref{denpert}) becomes
\begin{equation}
\label{grwthfeq1}
f'+f^2+\left(\frac{\dot{H}}{H^2}+2\right)f=\frac{3}{2}\frac{G_{eff}}{G}\omm,
\end{equation}
where $f'=df/d\ln a$. It is very interesting that the solution of
the equation can be approximated as $f=\omm^\gamma$
\cite{peebles,fry,lightman,silveira,wang,gong08b} and the growth
index $\gamma$ can be obtained for some general models. The
approximation was first proposed by Peebles for the matter dominated
universe as $f(z=0)=\Omega^{0.6}_0$ \cite{peebles}; then a more
accurate approximation, $f(z=0)=\Omega^{4/7}_0$, for the same model
was derived in \cite{fry,lightman}. For a dynamical dark energy
model with slowly varying $w$ and zero curvature, the approximation
$f(z)=\Omega(z)^\gamma$ was given in \cite{silveira,wang}. For more
general dynamical dark energy models in flat space, it was found
that $\gamma=0.55+0.05[1+w(z=1)]$ with $w>-1$ and
$\gamma=0.55+0.02[1+w(z=1)]$ with $w<-1$ \cite{linder05,linder07}.
For the flat DGP model, $\gamma=11/16$ \cite{linder07}.

Fitting $f(z)=\Omega(z)^\gamma$ to several sets of the most recently
observational data, lately one of us found that $\gamma  =
0.64^{+0.17}_{-0.15}$  for the  $\Lambda$CDM model and  $\gamma    =
0.55^{+0.14}_{-0.13}$ for the Dvali-Gabadadze-Porrati model
\cite{gong08b}. In this paper, we shall generalize such studies to
the case where the space is not flat. In particular, in Sec. II we
consider the dark energy model with constant $w$, while  in Sec. III
we discuss the DGP model. In Sec. IV, we apply the union compilation
of type Ia supernovae (SNe) data \cite{union}, the baryon acoustic
oscillation (BAO) measurement from the Sloan Digital Sky Survey
 \cite{sdss6}, the shift parameter, the acoustic scale $l_A$
and the redshift $z_*$ of the last scattering surface measured from
the Wilkinson Microwave Anisotropy Probe 5 yr data (WMAP5)
\cite{wmap5}, and the Hubble parameter data $H(z)$ \cite{hz1,hz2}.
We also use the growth factor data $f(z)$
\cite{porto,ness,guzzo,colless,tegmark,ross,angela,mcdonald,viel1,viel2}
to find the constraint on the growth index $\gamma$. We also
fit our simple analytical expression $f_{a}=\omm^\gamma+(\gamma-4/7)\omk$
to all the data sets by allowing the three parameters $\Omega_{m0}$,
$\Omega_{k0}$ and $\gamma$ all to vary. We conclude the paper in Sec. V.

It should be noted that when the Universe is not flat, the
approximation $f(z)=\Omega_m^{0.6}+\Omega_\Lambda/30$ was proposed
in \cite{martel} and
$f(z=0)=\Omega^{0.6}_{m0}+\Omega_{\Lambda0}(1+\Omega_{m0}/2)/70$ in
\cite{lahav} for the $\Lambda$CDM model. From these works one can
see that the approximation for the growth factor may not be
simply $\Omega_m^\gamma$ when the curvature is not zero, and that
the growth index $\gamma$ can still be used as the signature of
modified gravity and dark energy models.

\section{Dark Energy Models with Curvature}
For the curved dark energy model with a constant equation of state $w$,
we have
\begin{equation}
\label{wcdmhdot}
\frac{\dot H}{H^2}=\frac{1}{2}\omk-\frac{3}{2}[1+w(1-\omm-\omk)].
\end{equation}
The energy conservation equation tells us that
\begin{equation}
\label{wcdmwmder}
\omm'=3w\omm(1-\omm-\omk)-\omm\omk.
\end{equation}
Substituting Eqs. (\ref{wcdmhdot}) and (\ref{wcdmwmder}) into Eq. (\ref{grwthfeq1}),
we get
\begin{equation}
\label{wcdmfeq}
[3w\omm(1-\omm-\omk)-\omm\omk]\frac{df}{d\omm}+f^2+\left[\frac{1}{2}+\frac{1}{2}\omk
-\frac{3}{2}w(1-\omm-\omk)\right]f=\frac{3}{2}\omm.
\end{equation}
Plugging $f=\omm^\gamma$ into Eq. (\ref{wcdmfeq}), we get
\begin{equation}
\label{wcdmfeq1}
[3w(1-\omm-\omk)-\omk]\omm\ln\omm\frac{d\gamma}{d\omm}+\left(\gamma
-\frac{1}{2}\right)[3w(1-\omm-\omk)-\omk]+\omm^\gamma-\frac{3}{2}\omm^{1-\gamma}
+\frac{1}{2}=0.
\end{equation}
At high redshift, both $1-\omm$ and $\omk$ are small quantities and
they evolve differently. If we expand Eq. (\ref{wcdmfeq1}) around
$\omm=1$, then we also need to deal with $\omk\sim \omm\,a$. If
$\omk=0$, to the first order of $(1-\omm)$, we get \cite{wang,Discrepancy}
\begin{equation}
\label{wcdmr}
\gamma=\frac{3(1-w)}{5-6w}+\frac{3}{125}\frac{(1-w)(1-3w/2)}{(1-6w/5)^2(1-12w/5)}(1-\omm).
\end{equation}
If we take $\omk$ as a constant, then to the lowest order, we get
\begin{equation}
\label{wcdmr0}
(\gamma-1/2)(3w+1)\omk=0,
\end{equation}
so $\gamma=1/2$ which is not consistent with other results.
Therefore, we cannot take $\omk$ as a constant. If we think that
$\omk$ and $1-\omm$ are independently small quantities, then to the
first order of $1-\omm$ and $\omk$, we get
\begin{equation}
\label{wcdmr1}
\gamma_b=\frac{3(w-1)(1-\omm)-(3w+1)\omk}{(6w-5)(1-\omm)-2(3w+1)\omk}.
\end{equation}
If $\omk=0$,  Eq. (\ref{wcdmr1}) recovers the result given by Eq.
(\ref{wcdmr}) for the flat space. If there is no dark energy,
$\omk=1-\omm$, then Eq. (\ref{wcdmr1}) tells us that $\gamma=4/7$
which is consistent with the results obtained in
\cite{fry,lightman}. For the $\Lambda$CDM model, we get
$\gamma=[6(1-\omm)-2\omk]/[11(1-\omm)-4\omk]$. Equation
(\ref{wcdmr1}) is also consistent with Eq. (15) in
\cite{Mortonsonetal} giving a very similar expression when expanded in
terms of $\Omega_{de}=1-\Omega_{m}-\Omega_{k}$ and $\Omega_k$.
Although the result in Eq. (\ref{wcdmr1}) is consistent with other
results, the expression for $\gamma$ is perhaps too complicated to
be a constant; and we derive below a more practical expression for
$\gamma$ to distinguish between different models. This means that
the approximation $\omm^\gamma$ needs to be modified so that we can
get a simple number to distinguish different models. Let us recall
the approximation used for the curved model with a cosmological
constant. Martel used the approximation
$f=\omm^{0.6}+\Omega_{\Lambda}/30$ \cite{martel}. In \cite{lahav},
the approximation $f=\omm^{0.6}+\Omega_{\Lambda}(1+\omm/2)/70$ was
used. Following these ideas, we propose the approximation
$f_a=\omm^\gamma+\beta\omk$. Now $\omm$ and $\omk$ are independent
variables, so
\begin{equation}
\label{faderv}
f_a'=\beta\omk+\left(\beta\frac{\omk}{\omm}+\gamma\omm^{\gamma-1}\right)\omm'
+(\omm^\gamma+\omm^{\gamma-1}\omm')\omk\ln\omm\frac{\partial\gamma}{\partial\omk}
+\omm^\gamma\omm'\ln\omm\frac{\partial\gamma}{\partial\omm}
\end{equation}
Substituting the expression of $f_a$, Eqs. (\ref{wcdmwmder}) and
(\ref{faderv}) into Eq. (\ref{grwthfeq1}), expanding all quantities
around $\omm=1$ and $\omk=0$, and then keeping terms linear in $1-\omm-\omk$ and
$\omk$, we finally find that
\begin{equation}
\label{wcdmr4}
\gamma=\frac{3(1-w)}{5-6w},\quad \beta=\gamma-\frac{4}{7}.
\end{equation}
Therefore, the approximation becomes
\begin{equation}
\label{fapprox}
f_a=\omm^\gamma+(\gamma-4/7)\omk,\quad \gamma=\frac{3(1-w)}{5-6w}.
\end{equation}
For this approximation, we have only one constant $\gamma$ and this
growth index can be used as the fingerprint of the model. When
$\omk=0$, the approximation (\ref{fapprox}) recovers the familiar
result. For the $\Lambda$CDM model, the approximation is
$f_a=\omm^{6/11}-2\omk/77$. This result is consistent with Eq.
(15) in \cite{Mortonsonetal} for $\gamma$ when the latter is used
into their function $f(a)=\Omega_m(a)^{\gamma}$ and expanded around
$\Omega_k=0$.

\begin{figure}[htp]
\centering
\includegraphics[width=12cm]{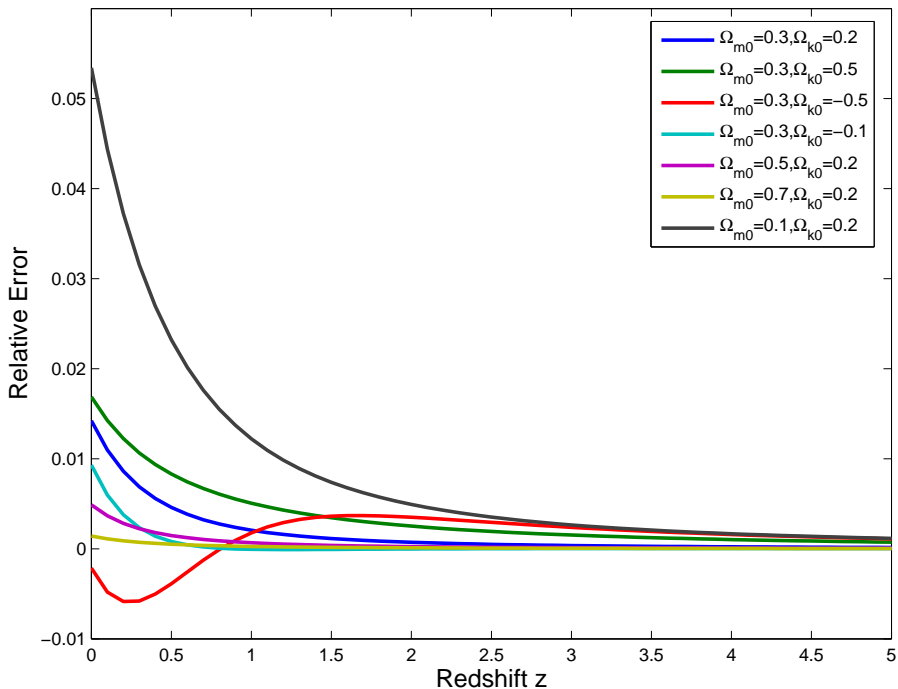}
\caption{The relative difference between the growth factor $f$ and the approximation
$f_a$ with $\gamma=6/11$ in Eq. (\ref{fapprox}) for the $\Lambda$CDM model.}
\label{lcdmferr}
\end{figure}

\begin{figure}[htp]
\centering
\includegraphics[width=12cm]{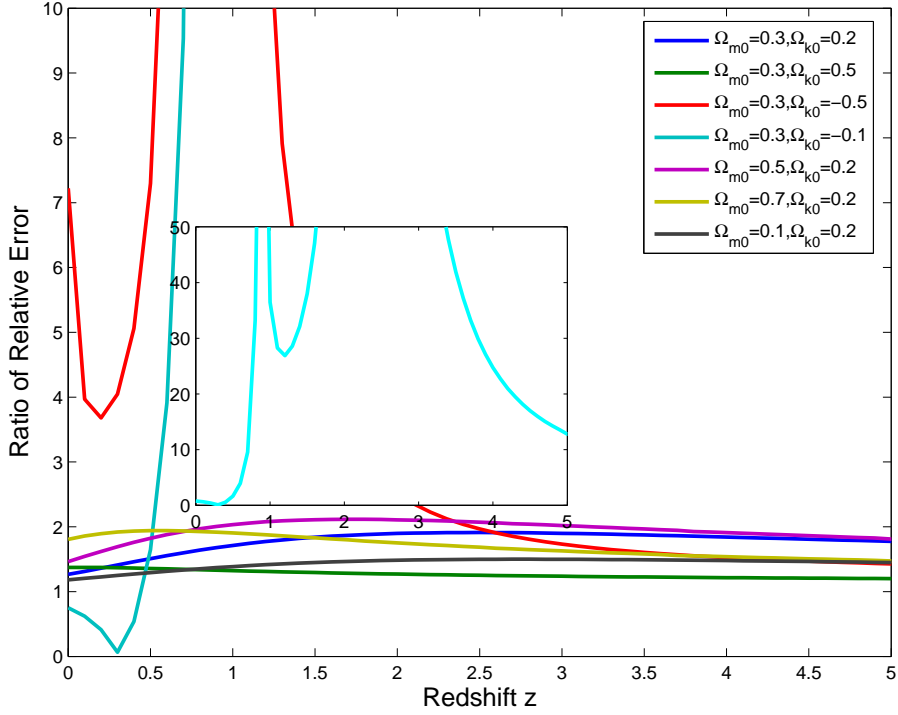}
\caption{The ratio $|(\omm^{\gamma_b}-f)/(f_a-f)|$ of the two different
approximations of the growth factor for the $\Lambda$CDM model.
Since the case with $\Omega_{m0}=0.3$ and $\Omega_{k0}=-0.1$ was cut
off in the plot, we show it explicitly in the inset.}
\label{lcdmfr}
\end{figure}

\begin{figure}[htp]
\centering
\includegraphics[width=12cm]{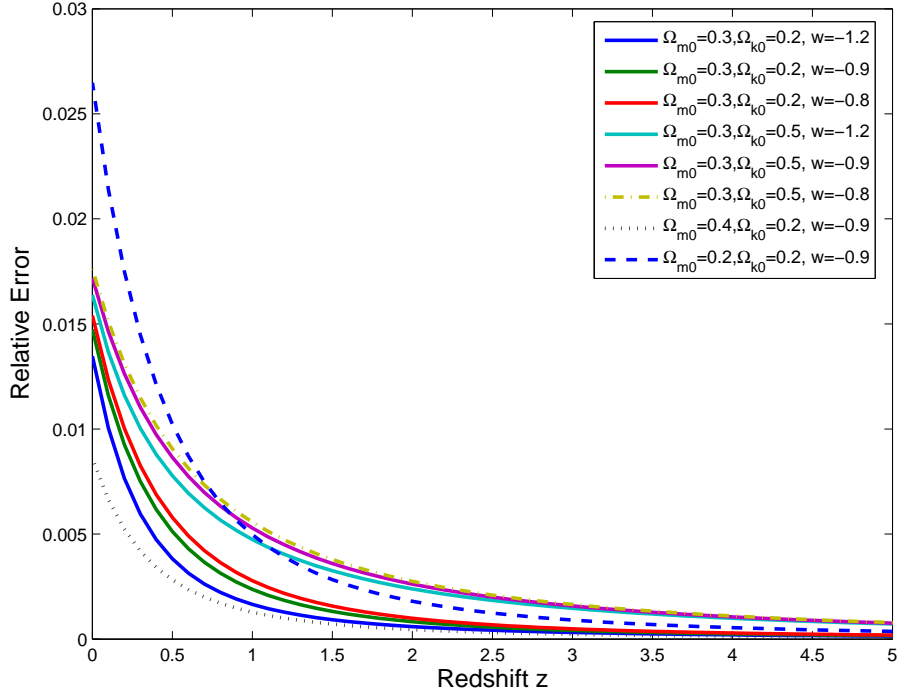}
\caption{The relative difference between the growth factor $f$ and the approximation
$f_a$ in Eq. (\ref{fapprox}) for the dark energy model with constant $w$.}
\label{wcdmferr}
\end{figure}

\begin{figure}[htp]
\centering
\includegraphics[width=12cm]{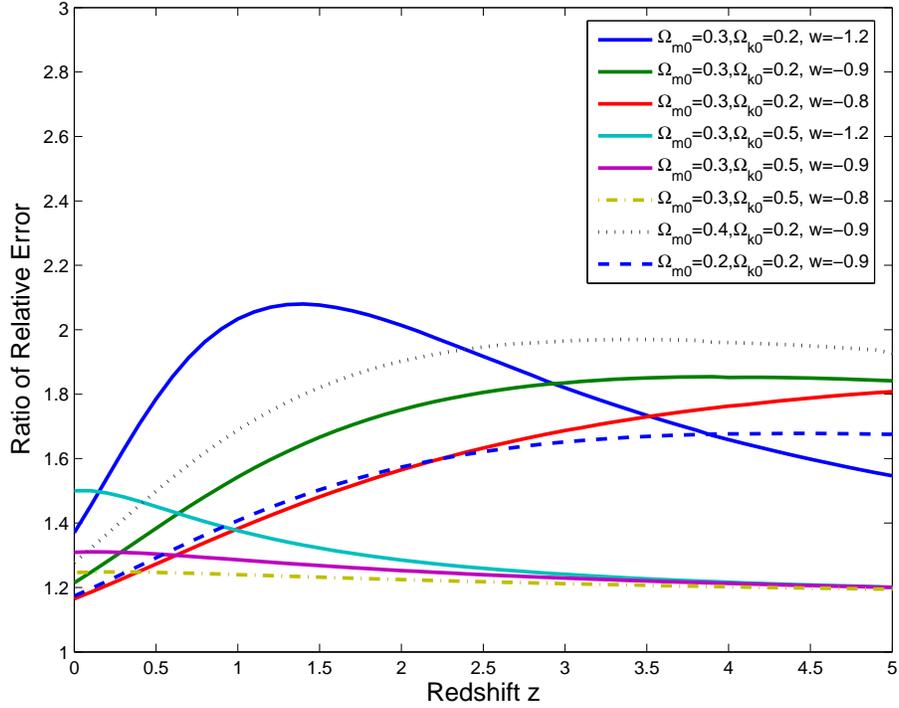}
\caption{The ratio $|(f_a-f)/(\omm^{\gamma_b}-f)|$ of the two
different approximations of the growth factor for
the dark energy model with constant $w$.}
\label{wcdmfr}
\end{figure}

In order to see how well the approximation $f_a$ fits the growth
factor $f$, we need to solve Eq. (\ref{grwthfeq1}) numerically with
the expression of $\omm$ and $\omk$. The dimensionless matter
density is
\begin{equation}
\label{wcdmwm}
\omm=\frac{\Omega_{m0} }{\Omega_{m0}+\Omega_{k0}(1+z)^{-1}
+(1-\Omega_{m0}-\Omega_{k0})(1+z)^{3w}},
\end{equation}
and the dimensionless curvature density is
\begin{equation}
\label{wcdmwk}
\omk=\frac{\Omega_{k0}(1+z)^{-1} }{\Omega_{m0}+\Omega_{k0}(1+z)^{-1}
+(1-\Omega_{m0}-\Omega_{k0})(1+z)^{3w}}.
\end{equation}
By using the above Eqs. (\ref{wcdmwm}) and (\ref{wcdmwk}), we solve
Eq. (\ref{grwthfeq1}) numerically to get the growth factor $f$ for
different values of $\Omega_{m0}$, $\Omega_{k0}$ and $w$. Then we
compare the approximation $f_a$ with $f$ by plotting the relative
error $(f_a-f)/f$ for the $\Lambda$CDM model in Fig. \ref{lcdmferr}.
From there we see that the error of the approximation is a few
percent, so $f_a$ approximates $f$ very well. We also compare the
approximation $\omm^{\gamma_b}$ with $\gamma_b$ given by Eq.
(\ref{wcdmr1}) with the approximation $f_a$ by plotting the absolute
value of the ratio $|(\omm^{\gamma_b}-f)/(f_a-f)|$ in
Fig.\ref{lcdmfr}, from which we see that $f_a$ approximates $f$
better. For the dark energy model with constant $w$, the accuracy of
the approximation $f_a$ is shown in Fig. \ref{wcdmferr} and the
comparison between the approximation of $f_a$ and $f_b$ is shown in
Fig. \ref{wcdmfr} for some typical values of $\Omega_{m0}$,
$\Omega_{k0}$ and $w$. The approximation $f_a$ is not only a better
approximation, but also has a simpler expression for the growth
index $\gamma$. Therefore, we should use Eq. (\ref{fapprox}) to
approximate the growth factor.

\section{DGP models with curvature}

For the DGP model, we have
\begin{equation}
\label{dgpgeff}
\frac{G_{\rm eff}}{G}=\frac{4\omm^2-4(1-\omk)^2+2\sqrt{1-\omk}
(3-4\omk+2\omm\omk+\omk^2)}{3\omm^2-3(1-\omk)^2
+2\sqrt{1-\omk}(3-4\omk+2\omm\omk+\omk^2)}.
\end{equation}
The Friedmann equation gives
\begin{equation}
\label{dgphdot}
\frac{\dot H}{H^2}=-\omk-\frac{3\omm(1-\omk)}{1+\omm-\omk}.
\end{equation}
The energy conservation equation tells us that
\begin{equation}
\label{dgpwmder}
\omm'=-\omm\left(3-2\omk-\frac{6\omm(1-\omk)}{1+\omm-\omk}\right).
\end{equation}
As discussed in the previous section, we consider the approximation
$f_a=\omm^\gamma+\beta\omk$. Substituting the expression of $f_a$,
Eqs. (\ref{dgpwmder}) and (\ref{faderv}) into Eq. (\ref{grwthfeq1}),
expanding all quantities around $\omm=1$ and $\omk=0$, and then keeping
terms linear in $1-\omm-\omk$ and $\omk$, we obtain
\begin{equation}
\label{dgpreq}
\left[2+\frac{7}{2}(\beta-\gamma)\right]\omk+\left[-4\gamma+\frac{11}{4}\right](1-\omm-\omk)=0.
\end{equation}
Therefore, the approximation of the growth factor for the DGP model
is
\begin{equation}
\label{dgpfa}
f_a=\omm^{11/16}+\left(\frac{11}{16}-\frac{4}{7}\right)\omk.
\end{equation}
On the other hand, if we approximate the growth factor $f$ by
$\omm^{\gamma_b}$, then set $\beta=0$ in Eq. (\ref{dgpreq}), finally we
find
\begin{equation}
\label{dgprb}
\gamma_b=\frac{11(1-\omm)-3\omk}{16(1-\omm)-2\omk}.
\end{equation}

\begin{figure}[htp]
\centering
\includegraphics[width=12cm]{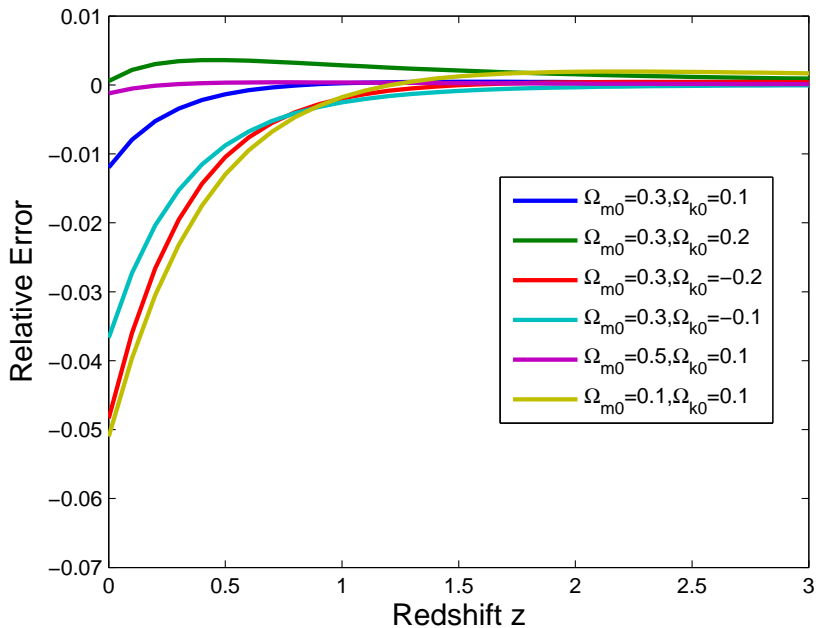}
\caption{The relative difference between the growth factor $f$ and $f_a$
with $\gamma=11/16$ for the DGP model.}
\label{dgpferr}
\end{figure}

In the DGP model, the dimensionless matter energy density is given
by
\begin{equation}
\label{dgpwm}
\omm=\frac{\Omega_{m0}(1+z)^3}{\Omega_{k0}(1+z)^2+[
\Omega_{r0}+\sqrt{\Omega_{m0} (1+z)^3+\Omega_{r0}^2}\,]^2},
\end{equation}
where
$\Omega_{r0}=(1-\Omega_{m0}-\Omega_{k0})/2\sqrt{1-\Omega_{k0}}$. The
dimensionless curvature energy density is
\begin{equation}
\label{dgpwk}
\omk=\frac{\Omega_{k0}(1+z)^2}{\Omega_{k0}(1+z)^2+[
\Omega_{r0}+\sqrt{\Omega_{m0} (1+z)^3+\Omega_{r0}^2}\,]^2}.
\end{equation}
Combining Eqs. (\ref{grwthfeq1}), (\ref{dgpgeff}), (\ref{dgphdot}), (\ref{dgpwm})
 and (\ref{dgpwk}), we get the evolution of the
growth factor $f$, and the result is then compared with the
approximation $f_a$. In Fig. \ref{dgpferr}, we show the accuracy of
the approximation by plotting $(f_a-f)/f$.

From Fig. \ref{dgpferr}, we see that the error is under 10\%. We
also checked the accuracy of the approximation $\Omega_m^{\gamma_b}$
and find that the approximation $f_a$ is usually better for the
model we are interested in.

\section{Observational Constraints}

Now we use the observational data to fit the dark energy model with
constant $w$ and the DGP model. The parameters in the models are
determined by minimizing
$\chi^2=\chi^2_{sn}+\chi^2_{bao}+\chi^2_{cmb}+\chi^2_h$. For the
Type Ia SNe data, we use the reduced union compilation of 307 Type
Ia SNe \cite{union}. The union compilation has 57 nearby Type Ia SNe
and 250 high-$z$ Type Ia SNe. It includes the Supernova Legacy
Survey \cite{astier} and the ESSENCE Survey \cite{riess,essence},
the older observed SNe data, and the extended data set of distant SNe
observed with the Hubble space telescope. To fit the Type Ia SNe
data, we define
\begin{equation}
\label{chi}
\chi^2_{sn}=\sum_{i=1}^{307}\frac{[\mu_{obs}(z_i)-\mu(z_i)]^2}{\sigma^2_i},
\end{equation}
where the extinction-corrected distance modulus
$\mu(z)=5\log_{10}[d_L(z)/{\rm Mpc}]+25$, $\mu_{obs}$ is the observed distance modulus, $\sigma_i$ is the total
uncertainty in the SNe data, and the luminosity distance is
\begin{equation}
\label{lum}
d_L(z)=\frac{1+z}{H_0\sqrt{|\Omega_{k}|}} {\rm
sinn}\left[\sqrt{|\Omega_{k}|}\int_0^z
\frac{dz'}{E(z')}\right],
\end{equation}
where
\begin{equation}
{\rm sinn}(\sqrt{|\Omega_k|}x)=\begin{cases}
\sin(\sqrt{|\Omega_k|}x),& {\rm if}\ \Omega_k<0,\\
\sqrt{|\Omega_k|}x, & {\rm if}\  \Omega_k=0, \\
\sinh(\sqrt{|\Omega_k|}x) & {\rm if}\  \Omega_k>0,
\end{cases}
\end{equation}
and the dimensionless Hubble parameter
$E(z)=H(z)/H_0=[\Omega_0(1+z)^3+(1-\Omega_0)(1+z)^{3(1+w)}]^{1/2}$
for the dark energy model with constant $w$ and
$E(z)=[\Omega_0(1+z)^3+(1-\Omega_0)^2/4]^{1/2}+(1-\Omega_0)/2$ for
the DGP model.

To use the BAO measurement from the Sloan Digital Sky Survey data, we define
\cite{sdss6}
\begin{equation}
\label{baochi2}
\chi^2_{bao}=\left(\frac{r_z(z_d)/D_V(z=0.2)-0.198}{0.0058}\right)^2
+\left(\frac{r_z(z_d)/D_V(z=0.35)-0.1094}{0.0033}\right)^2,
\end{equation}
where the effective distance is
\begin{equation}
\label{dvdef}
D_V(z)=\left[\frac{d_L^2(z)}{(1+z)^2}\frac{z}{H(z)}\right]^{1/3}.
\end{equation}
The redshift $z_d$ is fitted with the formulas \cite{hu98}
\begin{equation}
\label{zdfiteq}
z_d=\frac{1291(\Omega_m h^2)^{0.251}}{1+0.659(\Omega_m h^2)^{0.828}}[1+b_1(\Omega_b h^2)^{b_2}],
\end{equation}
\begin{equation}
\label{b1eq}
b_1=0.313(\Omega_m h^2)^{-0.419}[1+0.607(\Omega_m h^2)^{0.674}],\quad b_2=0.238(\Omega_m h^2)^{0.223},
\end{equation}
and the comoving sound horizon is
\begin{equation}
\label{rshordef}
r_s(z)=\int_0^\infty \frac{dz'}{c_s(z')E(z')},
\end{equation}
where the sound speed $c_s(z)=1/\sqrt{3[1+\bar{R_b}/(1+z)}]$,
$\bar{R_b}=315000\Omega_b h^2(T_{cmb}/2.7{\rm K})^{-4}$.

To implement the WMAP5 data, we need to add three fitting parameters
$R$, $l_a$ and $z_*$, so $\chi^2_{cmb}=\Delta x_i {\rm
Cov}^{-1}(x_i,x_j)\Delta x_j$, where $x_i=(R,\ l_a,\ z_*)$ denotes
the three parameters for the WMAP5 data, $\Delta x_i=x_i-x_i^{obs}$ and
Cov$(x_i,x_j)$ is the covariance matrix for the three parameters
\cite{wmap5}. The acoustic scale $l_A$ is
\begin{equation}
\label{ladefeq}
l_A=\frac{\pi d_L(z_*)}{(1+z_*)r_s(z_*)},
\end{equation}
where the redshift $z_*$ is given by \cite{hu96}
\begin{equation}
\label{zstareq}
z_*=1048[1+0.00124(\Omega_b h^2)^{-0.738}][1+g_1(\Omega_m h^2)^{g_2}]=1090.04\pm 0.93,
\end{equation}
\begin{equation}
g_1=\frac{0.0783(\Omega_b h^2)^{-0.238}}{1+39.5(\Omega_b h^2)^{0.763}},\quad
g_2=\frac{0.560}{1+21.1(\Omega_b h^2)^{1.81}}.
\end{equation}
The shift parameter
\begin{equation}
\label{shift}
R=\frac{\sqrt{\Omega_{m}}}{\sqrt{|\Omega_{k}|}}{\rm
sinn}\left(\sqrt{|\Omega_{k}|}\int_0^{z_*}\frac{dz}{E(z)}\right)=1.710\pm 0.019.
\end{equation}

Simon, Verde, and Jimenez obtained the Hubble parameter $H(z)$ at
nine different redshifts from the differential ages of passively
evolving galaxies \cite{hz1}. Recently, the authors in \cite{hz2}
obtained $H(z=0.24)=83.2\pm 2.1$ and $H(z=0.43)=90.3\pm 2.5$ by
taking the BAO scale as a standard ruler in the radial direction. To
use these 11 $H(z)$ data, we define
\begin{equation}
\label{hzchi}
\chi^2_h=\sum_{i=1}^{11}\frac{[H_{obs}(z_i)-H(z_i)]^2}{\sigma_{hi}^2},
\end{equation}
where $\sigma_{hi}$ is the $1\sigma$ uncertainty in the $H(z)$ data.
We also add the prior $H_0=72\pm 8$ km/s/Mpc given by Freedman {\it
et al.} \cite{freedman}. The likelihood for the parameters in the
model and the nuisance parameters $\Omega_b h^2$ and $H_0$ is
computed using a Monte Carlo Markov Chain (MCMC). The MCMC method
randomly chooses values for the above parameters, evaluates $\chi^2$,
and determines whether to accept or reject the set of parameters
using the Metropolis-Hastings algorithm. The set of parameters that
is accepted to the chain forms a new starting point for the next
process, and the process is repeated for a sufficient number of
steps until the required convergence is reached. Our MCMC code is
based on the publicly available package COSMOMC \cite{cosmomc}.

\begin{figure}[htp]
\centering
\includegraphics[width=12cm]{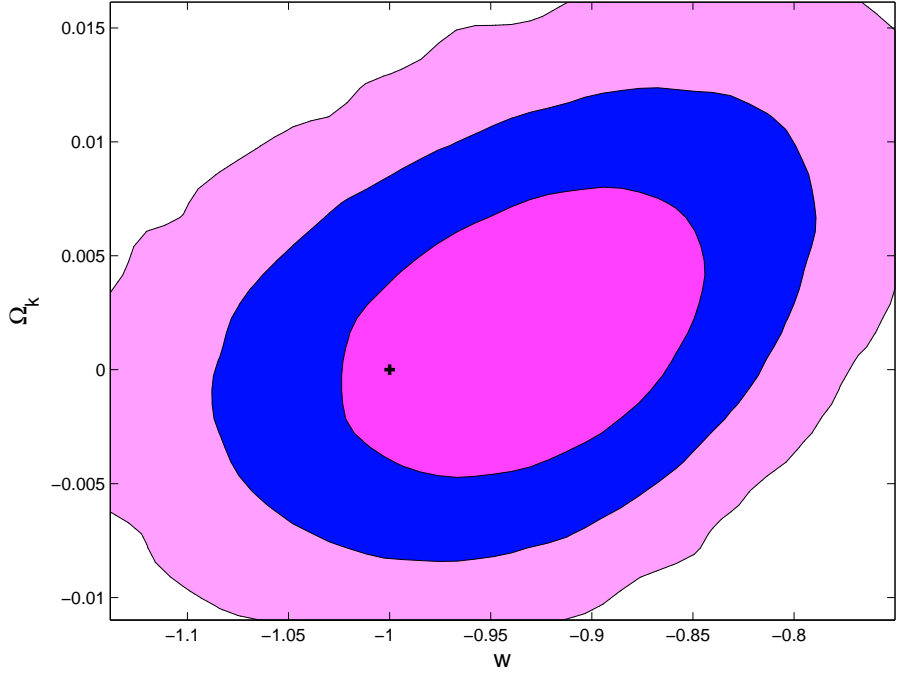}
\caption{The $1\sigma$ and $2\sigma$ contours of $\Omega$ and $w$ by fitting
the dark energy model with constant $w$ to the combined data. The point with $+$
denotes the flat $\Lambda$CDM point $\Omega_k=0$ and $w=-1$.}
\label{wcdmcont}
\end{figure}

\begin{figure}[htp]
\centering
\includegraphics[width=12cm]{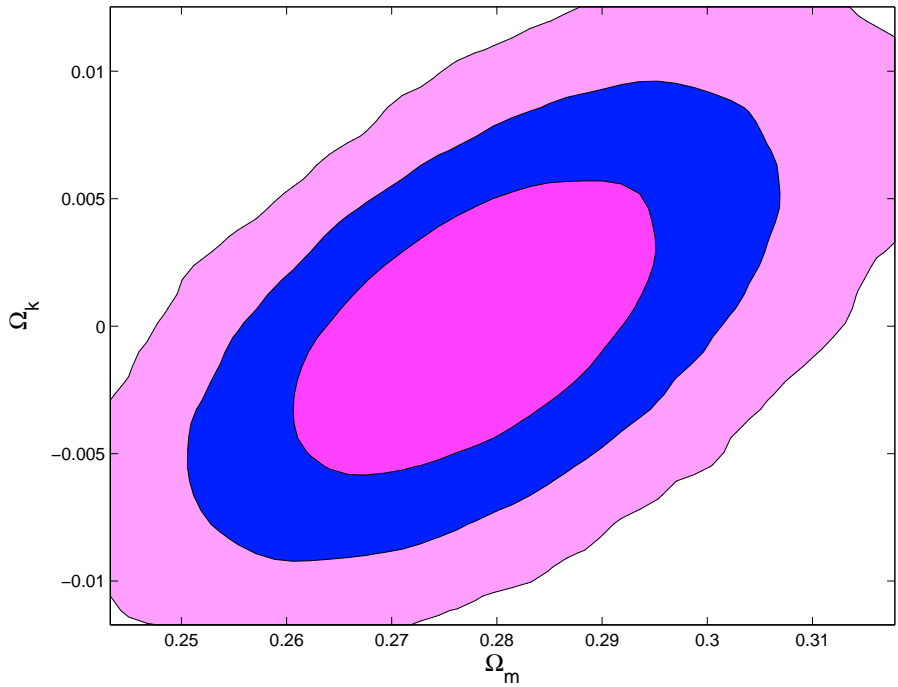}
\caption{The $1\sigma$ and $2\sigma$ contours of
$\Omega_m$ and $\Omega_k$ for the $\Lambda$CDM model.}
\label{lcdmcont}
\end{figure}

\begin{figure}[htp]
\centering
\includegraphics[width=12cm]{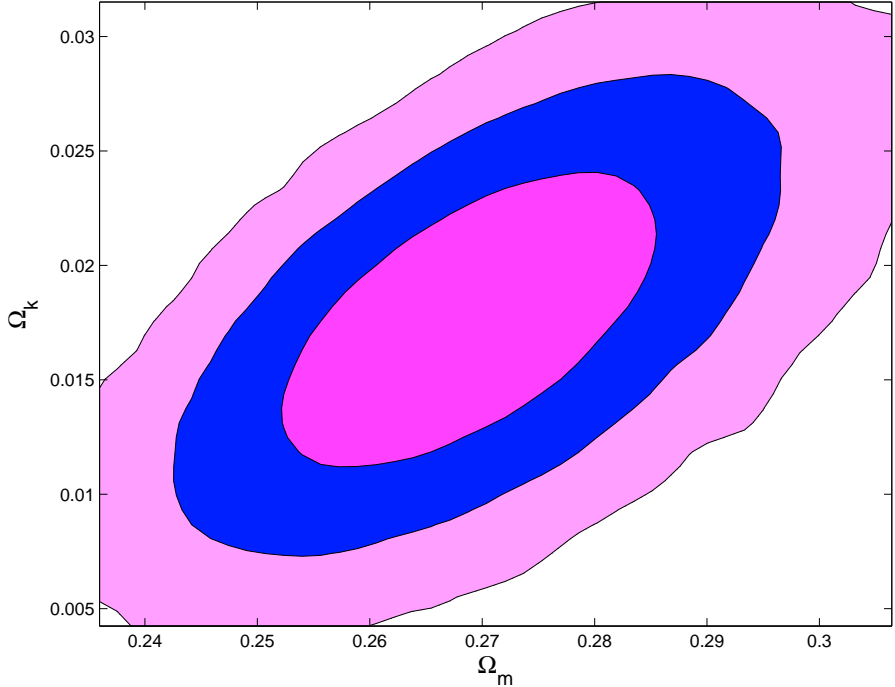}
\caption{The $1\sigma$ and $2\sigma$ contours of
$\Omega_m$ and $\Omega_k$ for the DGP model.}
\label{dgpcont}
\end{figure}

By fitting the dark energy model with constant $w$ to the combined
data (except the growth data that we use next), we get $\chi^2=332.6$,
$\Omega_{m0}=0.276^{+0.028}_{-0.026}$,
$\Omega_{k0}=0.002\pm 0.010$ and $w=-0.93^{+0.14}_{-0.15}$. The
$1\sigma$, $2\sigma$ and $3\sigma$ contours of $\Omega_{k0}$ and $w$
are shown in Fig. \ref{wcdmcont}. By fitting the $\Lambda$CDM model
to the combined data, we get $\chi^2=333.9$,
$\Omega_{m0}=0.277^{+0.025}_{-0.023}$ and $\Omega_{k0}=0.0002\pm
0.0081$. The $1\sigma$, $2\sigma$ and $3\sigma$ contours of
$\Omega_{m0}$ and $\Omega_{k0}$ are shown in Fig. \ref{lcdmcont}. By
fitting the DGP model to the combined data, we get $\chi^2=345.7$,
$\Omega_{m0}=0.268^{+0.024}_{-0.022}$ and $\Omega_{k0}=0.018\pm
0.009$. The $1\sigma$, $2\sigma$ and $3\sigma$ contours of
$\Omega_{k0}$ and $w$ are shown in Fig. \ref{dgpcont}.

Next, if we use constraints on $\Omega_{m0}$ and $\Omega_{k0}$ from
previous step and fit $\omm^\gamma+(\gamma-4/7)\omk$ to the growth
factor data $f(z)$ alone from \cite{porto,ness,guzzo,colless,tegmark,%
ross,angela,mcdonald,viel1,viel2},
we can get a constraint on the growth index $\gamma$. For the
curved $\Lambda$CDM model with the best fit value
$\Omega_{m0}=0.277$ and $\Omega_{k0}=0.0002$, we find that
$\chi^2=4.54$ and $\gamma_\Lambda=0.65^{+0.17}_{-0.15}$. The
theoretical value $\gamma_\infty=6/11=0.55$ is consistent with the
observation at the $1\sigma$ level. For the DGP model with the best
fit value $\Omega_{m0}=0.268$ and $\Omega_{k0}=0.018$, we find
$\chi^2=5.79$ and $\gamma_{DGP}=0.53^{+0.14}_{-0.12}$ which is not
consistent with the theoretical value $\gamma_\infty=11/16=0.6875$
at the $1\sigma$ level.

We also fit $\omm^\gamma+(\gamma-4/7)\omk$ to the combined Type Ia,
the BAO, the CMB, the Hubble parameter $H(z)$, and the growth factor $f(z)$ data, and allowing for
the three parameters to vary. We obtain the following
constraints on the parameters $\Omega_{m0}$, $\Omega_{k0}$
and $\gamma$. For the curved $\Lambda$CDM model,
we find that $\chi^2=338.4$, $\Omega_{m0}=0.276^{+0.03}_{-0.024}(1\sigma)^{+0.041}_{-0.033}(2\sigma)$,
$\Omega_{k0}=-0.0002^{+0.0095}_{-0.0086}(1\sigma)^{+0.0132}_{-0.0118}(2\sigma)$, and
$\gamma=0.64^{+0.48}_{-0.32}(1\sigma)^{+0.73}_{-0.43}(2\sigma)$. For the DGP model, we find that
$\chi^2=351.47$, $\Omega_{m0}=0.268^{+0.028}_{-0.024}(1\sigma)^{+0.039}_{-0.033}(2\sigma)$,
$\Omega_{k0}=0.017^{+0.011}_{-0.009}(1\sigma)^{+0.015}_{-0.013}(2\sigma)$, and
$\gamma=0.52^{+0.40}_{-0.27}(1\sigma)^{+0.61}_{-0.35}(2\sigma)$. Proceeding in this second
way shows that current data that we used is not enough in order to put conclusive constraints
on $\gamma$ and the underlying theory of gravity; however, this is likely
to change with future data from future high precision missions such as  PLANCK, JDEM, LSST, and others.

\section{Discussions}

We find that the simple analytical expression
$f_{a}=\omm^\gamma+(\gamma-4/7)\omk$ can be used to approximate very well
the growth factor $f$ when the curvature is not zero and this is the main result
of the paper. The curvature parameter is included, as shown above, into
the function $f_a$ which is found to provide a better fit to the growth
factor than $\omm^{\gamma}$ (with curvature included in $\gamma$) when compared
to the growth factor evaluated numerically from the differential
equation (this is shown in Figs 3-5). We also find that,
with $f_{a}$ as given above, the asymptotic value of $\gamma$ is the
same as that in flat space and it provides distinctive information about
a dark energy model versus a modification of gravity. For example,
$\gamma\approx 0.55$ for the $\Lambda$CDM model and $\gamma=0.6875$
for the DGP model.

We explored then some constraints on $\gamma$ using the form of $f_a$
above and current data from Type Ia SNe, BAO, WMAP5, $H(z)$ data, and growth
factor data. Unlike previous analyses which used the shift parameter for the
WMAP5 data and the $A$ parameter for the BAO data \cite{gong08b},
here we used the shift parameter $R$, the acoustic scale $l_A$, the
redshift $z_*$ and their covariance matrix for the WMAP5 data
\cite{wmap5}, and the parameter $r_s(z_d)/D_V(z)$ at two different
redshifts for the BAO data.

First, we fit the models to the combined
Type Ia SNe, BAO, WMAP5, and $H(z)$ data and obtained
$\Omega_{m0}=0.277^{+0.025}_{-0.023}$, $\Omega_{k0}=0.0002\pm
0.0081$ and $\chi^2=333.9$ for the curved $\Lambda$CDM model, and
$\Omega_{m0}=0.268^{+0.024}_{-0.022}$, $\Omega_{k0}=0.018\pm 0.009$
and $\chi^2=345.7$ for the DGP model. The DGP model is found strongly
disfavored by the data, a result consistent with other analysis
\cite{song,song05,roy06,gong,zhu,gong08b}. Comparing with the WMAP5
result \cite{wmap5}, we see that the error bar becomes a little
bigger with the addition of the $H(z)$ data. The reason is due to
the larger uncertainties in the $H(z)$ data. Now, using the best fit
values for $\Omega_{m0}$ and $\Omega_{k0}$ from the previous step and
the growth data of $f(z)$, we obtained constraints on the growth index $\gamma$.
For the curved $\Lambda$CDM model, we find that
$\gamma_\Lambda=0.65^{+0.17}_{-0.15}$, which is consistent with the
theoretical value 0.55. This result is also consistent with those in
\cite{ness,porto,gong08b}. For the DGP model, we find that
$\gamma_{DGP}=0.53^{+0.14}_{-0.12}$. The theoretical value
$\gamma=0.6875$ lies outside the $1\sigma$ bounds thus again disfavoring
the DGP model.

Next, we fit $f_a$ to all the data sets by allowing the three parameters $\Omega_{m0}$, $\Omega_{k0}$
and $\gamma$ to vary. For the curved $\Lambda$CDM model,
we find that $\chi^2=338.4$, $\Omega_{m0}=0.276^{+0.03}_{-0.024}(1\sigma)^{+0.041}_{-0.033}(2\sigma)$,
$\Omega_{k0}=-0.0002^{+0.0095}_{-0.0086}(1\sigma)^{+0.0132}_{-0.0118}(2\sigma)$, and
$\gamma=0.64^{+0.48}_{-0.32}(1\sigma)^{+0.73}_{-0.43}(2\sigma)$. For the curved DGP model, we find that
$\chi^2=351.47$, $\Omega_{m0}=0.268^{+0.028}_{-0.024}(1\sigma)^{+0.039}_{-0.033}(2\sigma)$,
$\Omega_{k0}=0.017^{+0.011}_{-0.009}(1\sigma)^{+0.015}_{-0.013}(2\sigma)$, and
$\gamma=0.52^{+0.40}_{-0.27}(1\sigma)^{+0.61}_{-0.35}(2\sigma)$. Thus, the current data
is still not sufficient for us to obtain conclusive constraints, when we fit all these three parameters
simultaneously. But this is likely to be improved substantially with future high precision missions.

Finally, the inclusion of the curvature in future analyses of cosmological data in order to constrain
the origin of cosmic acceleration was recently discussed and stressed in the literature, e.g.
\cite{wmap5,virey,gong08}.
This includes the question of distinguishing   dark energy from modified gravity using
the growth data in the presence of spatial curvature. We find here that the analytical expression
$f_{a}=\omm^\gamma+(\gamma-4/7)\omk$ approximates very well the growth factor $f$ when curvature
is added and should be useful for such future analyses.
\begin{acknowledgments}
Y.G. thanks the hospitality of Baylor University where this work was
completed and the support by NNSFC under Grant No. 10605042. M.I.
acknowledges partial support from the Hoblitzelle foundation and the
Texas Space Grant Consortium.
\end{acknowledgments}


\begin{thebibliography}{sp99}
\bibitem{Discrepancy} We find a small discrepancy in the denominator
of the second term of equation (\ref{wcdmr}) compared to the widely
cited expression for the growth, i.e. equation (B11) from pioneering
work by \cite{wang}.
\end{thebibliography}
\end{document}